\documentclass[12pt]{article}

\usepackage{newtxtext,newtxmath}

\usepackage{graphicx}

\usepackage[letterpaper,margin=1in]{geometry}

\renewenvironment{abstract}
	{\quotation}
	{\endquotation}

\date{}

\makeatletter
\renewcommand{\fnum@figure}{\textbf{Figure \thefigure}}
\renewcommand{\fnum@table}{\textbf{Table \thetable}}
\makeatother

\usepackage{scicite}

\usepackage{url}

\newcommand{\textalpha}{\ensuremath{\alpha}}
\newcommand{\coloneqq}{:=}
\newcommand{\eqqcolon}{=:}
\newcommand{\Tr}{\mathrm{Tr}}
\newcommand{\im}{\mathrm{i}}
\newcommand{\enquote}[1]{{``#1''}}

\newcommand{\eff}{\mathrm{eff}}         
\newcommand{\mf}{\mathrm{mf}}           
\newcommand{\loc}{\mathrm{loc}}         
\newcommand{\Ph}{\mathrm{P}}            
\newcommand{\Qu}{\mathrm{Q}}            

\newcommand{\mb}{\mathbf}
\newcommand{\dagg}{\dagger}

\newcommand{\op}[1]{\mb{#1}}

\newcommand{\dul}[1]{\underline{\underline{#1}}}

\newcommand{\oH}{\op{H}}

\newcommand{\oU}{\op{U}}
\newcommand{\oUd}{\op{U}^{\dagg}}

\newcommand{\dira}{\mu}
\newcommand{\dirb}{\nu}
\newcommand{\oS}{\op{S}}
\newcommand{\Sx}{\oS^{x}}
\newcommand{\Sy}{\oS^{y}}
\newcommand{\Sz}{\oS^{z}}
\newcommand{\Sp}{\oS^{+}}
\newcommand{\Sm}{\oS^{-}}
\newcommand{\Sa}{\oS^{\dira}}
\newcommand{\Sb}{\oS^{\dirb}}

\newcommand{\Iz}{\op{I}^{z}}
\newcommand{\Ia}{\op{I}^{\dira}}

\newcommand{\Sai}{\Sa_{i}}

\newcommand{\Sbj}{\Sb_{j}}

\newcommand{\vS}{\vec{\op{S}}}

\newcommand{\vI}{\vec{\op{I}}}

\newcommand{\oV}{\op{V}}

\newcommand{\vV}{\vec{\oV}}

\newcommand{\cVz}{V^{z}}
\newcommand{\cVa}{V^{\dira}}
\newcommand{\cVb}{V^{\dirb}}

\newcommand{\cvV}{\vec{V}}

\newcommand{\JQ}{J_{\Qu}}

\newcommand{\zeff}{z_{\eff}}

\newcommand{\TSD}{T_{\text{SD}}}
\newcommand{\TZQT}{T_{\text{ZQ}}}
\newcommand{\GZQT}{\Gamma_{\text{ZQ}}}

\newcommand{\expval}[1]{\langle#1\rangle}
\newcommand{\mfav}[1]{\overline{#1}^{\mf}}

\newcommand{\refcite}[1]{Ref.~\cite{#1}} 
\newcommand{\refsec}[1]{Section~\ref{#1}} 
\newcommand{\refsupsec}[1]{section~\ref{#1}} 
\newcommand{\refeqn}[1]{Equation~\eqref{#1}} 
\newcommand{\refsupeqn}[1]{equation~\eqref{#1}} 
\newcommand{\reffig}[1]{Figure~\ref{#1}} 
\newcommand{\refsupfig}[1]{figure~\ref{#1}} 
\newcommand{\reftab}[1]{Table~\ref{#1}} 
\newcommand{\refsuptab}[1]{table~\ref{#1}} 

\newcommand{\SI}[2]{#1~#2} 
\newcommand{\hertz}{\text{Hz}}
\newcommand{\degree}{\!\text{$^{\circ}$}}
\newcommand{\millisecond}{\text{ms}}
\newcommand{\microsecond}{\text{\textmu s}}

\newcommand{\zqop}{\op{S}_{12}^{\text{ZQ}}}

\def\scititle{
	First-principles simulation of spin diffusion in static solids using dynamic mean-field theory
}
\title{\bfseries \boldmath \scititle}

\author{
	Timo~Gräßer$^{1\ast}$,
	G\"otz~S.~Uhrig$^{2}$,
	Matthias~Ernst$^{1\ast}$\and
	\small$^{1}$Institute for Molecular Physical Sciences, ETH Zürich, 8093 Zürich, Switzerland.\and 
	\small$^{2}$Condensed Matter Theory, Department of Physics, TU Dortmund University, 44221 Dortmund, Germany.\and 
	\small$^\ast$Corresponding authors. Emails: timo.graesser@phys.chem.ethz.ch, maer@ethz.ch
}

\begin{document} 

\maketitle

\begin{abstract} \bfseries \boldmath
The dynamics of disordered nuclear spin ensembles are the subject of nuclear magnetic resonance studies. Due to the through-space long-range dipolar interaction generically many spins are involved in the time evolution, so that exact brute force calculations are impossible. The recently established spin dynamic mean-field theory (spinDMFT) represents an efficient and unbiased alternative to overcome this challenge. The approach only requires the dipolar couplings as input and the only prerequisite for its applicability is that each spin interacts with a large number of other spins. In this article, we show that spinDMFT can be used to describe spectral spin diffusion in static samples and to simulate zero-quantum line shapes which eluded an efficient quantitative simulation so far to the best of our knowledge. We perform benchmarks for two test substances that establish an excellent match with published experimental data. As spinDMFT combines low computational effort with high accuracy, we suggest to use it for large-scale simulations of spin diffusion, which are important in various areas of magnetic resonance.
\end{abstract}

\newpage

\section{Introduction}
\noindent

The understanding of magnetic resonance phenomena is important in various research areas such as quantum computing \cite{divin00,raman04}, the study of nitrogen-vacancy centers \cite{rezai25,nagur26} as well as the broad fields of electron spin resonance and nuclear magnetic resonance (NMR) \cite{Abragam:1961vga,ernst90}. In many setups, spin dynamics is induced by the mutual dipolar interaction, which does not only lead to dephasing but also to polarization transfer. Under the right circumstances, this transfer occurs under conservation of the total spin polarization, which is usually called \enquote{spin diffusion} in the context of NMR. This term was originally coined by Bloembergen in 1949 \cite{Bloembergen.1949} to describe the polarization exchange in a strongly-coupled many-spin system that, under certain conditions, appears to follow a spatial diffusion equation. Today, spin diffusion is often used in a broader sense, describing dipolar-coupling based polarization-transfer processes even if they are not characterized by a diffusive behavior. However, the time evolution of the density operator under a static Hamiltonian is always coherent and can, in principle, be reversed, as has also been shown experimentally for the spin-diffusion process \cite{Zhang.1992}. Two forms of spin diffusion are often distinguished: (i) spatial spin diffusion between equivalent spins in space due to spatial differences in polarization levels. (ii) Spectral spin diffusion between spins with separated resonance frequencies due to non-equal initial polarization values of the different spins \cite{suter85,kubo88,Henrichs.1986}. In reality, one often has a superposition of the two forms of spin diffusion.

Spectral spin diffusion can be easily measured as long as the two lines are spectrally resolved \cite{suter85,Henrichs.1986,kubo88}. Measuring spatial spin diffusion is more complex due to the fact that all spins contribute to the same spectral line. One possibility is to generate local polarization wells using cross polarization from rare spins \cite{Zhang.1993} and to measure the decay of the localized polarization. A more direct measurement of spin diffusion requires strong pulsed gradients \cite{Zhang.1993} or the use of magnetic-resonance force microscopy detected NMR \cite{Eberhardt.2007}. 
  
A direct quantum-mechanical simulation of spin diffusion is impossible due to the large number of spins required to obtain a good agreement with reality. Simulations using classical equations of motion (Bloch equations) are successful for free-induction decays (FIDs) in three-dimensional lattices \cite{elsay15,stark18}, but may not be accurate for local quantum-mechanical processes such as spectral spin diffusion. Mostly, spin diffusion has been simulated using a perturbation approach (Fermi's Golden Rule) where the active coupling is the perturbation and the energy-level splitting of the zero-quantum transition is characterized by the zero-quantum line shape \cite{suter85,Henrichs.1986,kubo88,Ernst.1998nti}. While the magnitude of the orientation-dependent dipolar coupling is easily accessible from the distance between the two spins, the zero-quantum line shape is difficult to obtain theoretically and experimentally. Often, it is calculated from the convolution of the two single-quantum line shapes, which assumes that the broadening mechanisms of the two spectral lines are uncorrelated. Under magic-angle spinning (MAS), the zero-quantum line shape can be simulated using a limited number of spins \cite{dumez11,vesht11} and, therefore, a calculation of the spin-diffusion rate constant using the perturbation approach \cite{Kubo:1988wq} is feasible. Under MAS, even a direct simulation of spin diffusion is possible \cite{dumez12} using state-space restrictions \cite{Kuprov:2007ga,Kuprov:2008fu} to reduce the dimensionality of the Liouville space. However, attempts to implement such approaches in static solids failed to the best of our knowledge.

So far, an efficient, accurate, and unbiased first-principles approach to simulate spin diffusion in static solids has not been described. This is where our recently developed spin dynamic mean-field theory (spinDMFT) \cite{graes21} comes into play. The objective of this paper is to illustrate that spinDMFT is perfectly suited to fill the gap, enabling unbiased and well-justified first-principles calculations of spectral spin diffusion. In addition, the zero-quantum line shape can be simulated without additional assumptions such as uncorrelated single-quantum line shapes.

SpinDMFT was developed for large dense spin ensembles at high temperature. In this context, \enquote{dense} means that each spin has a large number of interaction partners. The term \enquote{high temperature} refers to the thermally available energy being much larger than all internal energy scales such as spin-spin or spin-field couplings; this entails that the initial density matrix is approximately proportional to the identity. The local spin dynamics is captured by an effective single-site problem with a time-dependent Gaussian mean-field and a closed self-consistency equation, which is solved numerically by iteration. The key control parameter is the number of interaction partners for each spin. If this number tends to infinity, each spin only experiences a classical field because in this limit the relative importance of non-commutativity vanishes \cite{stane14b}. In addition, the local spin environment consists of many contributions, which are only weakly correlated, so that the effective field behaves like a Gaussian-distributed random variable. 

SpinDMFT has been shown to successfully describe some standard NMR measurements, such as FIDs and spin echoes \cite{graes24} and is also capable of describing quadrupolar interactions \cite{graes26a}. In this article, we extend the spinDMFT approach to capture spectral spin diffusion in static solids and compare the approach with published experimental data for selectively \textsuperscript{13}C-labeled malonic acid and for dipotassium \textalpha-D-glucopyranose-1-phosphate dihydrate (GLP). We are well aware that solid-state NMR is mostly performed under magic-angle spinning (MAS), which is not considered in this study. The goal of this article is to highlight the enormous potential of spinDMFT on the example of static spectral spin diffusion, which is a long-standing theoretical problem. More relevant NMR experiments may be considered in future works after spinDMFT has been extended to more complex scenarios, such as MAS.

The article is structured as follows. First, we present the theoretical formulation of spinDMFT in \refsec{subsec:results:singlesite} and explain how it is systematically extended to capture spectral spin diffusion in \refsec{subsec:results:spindiff}. Subsequently, we present the simulation results and compare them to published experimental data in Sections~\ref{subsec:results:malnac} and~\ref{subsec:results:GLP}. A discussion and an outlook is provided in \refsec{sec:discussion}. Finally, we give an overview of the considered test substances in \refsec{sec:materials}. We explain how the crystal structure and the relevant dipolar couplings and chemical shifts are obtained.

\section{Results}
\label{sec:results}

\subsection{Definitions and single-site spinDMFT}
\label{subsec:results:singlesite}

\begin{figure}
    \centering
    \includegraphics[width=0.9\textwidth]{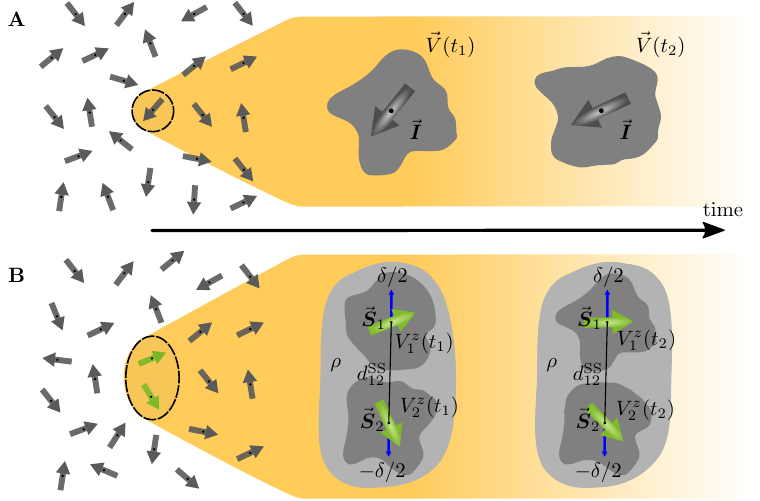}
    \caption{\textbf{Sketch of the effective models resulting from spinDMFT.} (\textbf{A}) SpinDMFT reduces a homogeneous spin lattice with many interacting spins (left) to a single spin interacting with a time dependent Gaussian mean-field (right). As the single spin changes its orientation over time, so does the mean-field, which is indicated here by the gray cloud. (\textbf{B}) SpinDMFT reduces the full lattice problem for spectral spin diffusion (left) to an effective model with explicit time-dependence (right). The coupling between the low-$\gamma$ spins (indicated by green arrows) is crucial and needs to be treated quantum-mechanically to capture spectral spin diffusion correctly. Therefore, the effective model treats both low-$\gamma$ spins explicitly. The dark gray clouds represent the mean-fields resulting from the high-$\gamma$ bath spins (indicated by gray arrows). The mean-fields are correlated, which is indicated by the light gray cloud. The blue arrows represent the chemical shifts acting on the low-$\gamma$ spins.}
    \label{fig:model}
\end{figure}

We consider a homogeneous spin system with $I=1/2$ at high temperatures, i.e., without any initial order. The spins interact according to the secular dipolar Hamiltonian truncated by a large external static magnetic field
\begin{align}
    \oH &= \frac12 \sum_{i,j} d^{\mathrm{II}}_{ij} \left( 3 \Iz_i \Iz_j - \vI_i \cdot \vI_j \right),
\end{align}
with 
\begin{align}
    d^{\mathrm{II}}_{ij} &\coloneqq \frac{1 - 3 \cos^2(\vartheta_{ij})}{2} \frac{\mu_0}{4\pi} \frac{\gamma_{\mathrm{I}}^2\hbar}{|\vec{r}_{ij}|^3}, & i&\neq j, 
\end{align}
where $\vec{r}_{ij}\coloneqq\vec{r}_{j}-\vec{r}_{i}$ is the distance vector between spins $i$ and $j$ and $\vartheta_{ij}$ is the angle between the internuclear vector and the static magnetic field $\vec{B}$. We set $d^{\mathrm{II}}_{ii}=0$ to rule out any self-interactions. The superscript of the coupling indicates the two species of spins that are coupled. The Hamiltonian can be rewritten as 
\begin{align}
    \oH &= \frac12 \sum_{i} \vI_i \cdot \vV_i
\end{align}
defining the local-environment operators
\begin{align}
    \vV_i &\coloneqq \sum_{j} d_{ij}^{\mathrm{II}} \dul{D} \, \vI_j, & \dul{D} &= 
    \begin{pmatrix}
        -1 & 0  & 0 \\
        0  & -1 & 0 \\
        0  & 0  & 2 \\
    \end{pmatrix}.
    \label{eqn:locenvfields}
\end{align}
The systematic control parameter of spinDMFT is the inverse of the number of interaction partners. In case of isotropic nearest-neighbor interactions, this number is simply the coordination number, i.e., the number of nearest neighbor spins. In case of long-range (e.g. dipolar) interactions, the number of interaction partners can be quantified by the effective coordination number
\begin{align}
    \zeff \coloneqq \frac{\left( \sum_{j}\left(d^{\mathrm{II}}_{ij}\right)^2 \right)^2}{ \sum_{j}\left(d^{\mathrm{II}}_{ij}\right)^4 },
\end{align}
which is independent of $i$ as the system is considered homogeneous. If $\zeff$ is sufficiently large, it is well-justified to replace each local-environment field by a dynamic Gaussian mean-field $\cvV(t)$. This leads to an effective single-site problem described by the mean-field Hamiltonian
\begin{align}
    \oH^{\mf}(t) &= \cvV(t) \cdot \vI,
    \label{eqn:mfHam}
\end{align}
which is sketched in \reffig{fig:model} A. Since we consider the high-temperature limit, the mean-field is zero on average, i.e., 
\begin{align}
    \mfav{ \cVa(t) } &= 0, & \mu&\in\{x,y,z\}.
    \label{eqn:selfcons_average}
\end{align}
Its second moments can be linked to the spin autocorrelations resulting in the self-consistency condition 
\begin{align}
    \mfav{ \cVa(t) \cVb(0) } &= \left(\JQ^{\mathrm{II}}\right)^2 \delta^{\dira\dirb} \left(D^{\dira\dira}\right)^2 \expval{ \Ia(t) \Ia(0) }, & \mu,\nu&\in\{x,y,z\},
    \label{eqn:selfcons}
\end{align}
where $D^{\dira\dira}$ are diagonal matrix elements of $\dul{D}$ and we defined the quadratic coupling constant
\begin{align}
    \left(\JQ^{\mathrm{II}}\right)^2 &= \sum_{j} \left(d^{\mathrm{II}}_{ij}\right)^2.
\end{align}
To compute the spin autocorrelations in the framework of spinDMFT, one needs to perform a classical average over the mean-field degrees of freedom as well as a quantum average over the remaining spin degrees of freedom, i.e., the single site. The formal equation for this reads 
\begin{align}
    \mfav{\expval{ \Ia(t) \Ia(0) }} &= \int \mathrm{D}\mathcal{V} \, p(\mathcal{V}) \,\expval{ \Ia(t)[\mathcal{V}] \Ia(0) }.
    \label{eqn:mfspinautocorr}
\end{align}
Here, $\mathcal{V}$ is a single time series of the mean-field and $p(\mathcal{V})$ is the mean-field's multivariate Gaussian probability distribution. The expectation value $\expval{ \Ia(t)[\mathcal{V}] \Ia(0) }$ is computed for the local site considering the time evolution generated by the Hamiltonian in \refeqn{eqn:mfHam} for a specific time series $\mathcal{V}$. The density operator is proportional to the identity in all considered expectation values in accordance with the considered high-temperature limit.

The defined self-consistency problem is closed and can be solved numerically by iteration. Starting from an initial guess for the spin autocorrelations, one computes the second mean-field moments via \refeqn{eqn:selfcons} and uses them to update the spin autocorrelations by means of \refeqn{eqn:mfspinautocorr}. This process is repeated until the spin autocorrelations converge, which typically requires only about 5 iteration steps. In practice, the path integral in \refeqn{eqn:mfspinautocorr} is evaluated by discretizing the time and applying a Monte-Carlo simulation. This entails an efficient importance sampling of the probability space by drawing a large set of mean-field time series according to the distribution $p(\mathcal{V})$. For each time series, one has to evolve a single spin under the specific time evolution of the dynamic mean-field. An average over the samples yields an estimate of the path integral up to a small statistical error. The simulation requires only a few minutes of computation time for typical numerical parameters. For more details on the derivation and numerical implementation of spinDMFT, we refer the reader to the original spinDMFT article \cite{graes21}. An open-source code collection for spinDMFT was published under \refcite{graes25}.

In principle, spinDMFT can be applied in a straightforward manner to inhomogeneous systems as long as the coordination number is large at all sites. However, if the quadratic coupling constant $\JQ^{\mathrm{II}}$ varies strongly from site to site, the self-consistency condition in \refeqn{eqn:selfcons} must be adapted to superpose several autocorrelations each representing a different site. This leads to a nested self-consistency problem as described in \refsupsec{sup:nestedselfcons}. If $\JQ^{\mathrm{II}}$ does not vary strongly, it is well-justified to consider single-site spinDMFT with an averaged quadratic coupling
\begin{align}
    J^{\mathrm{II}}_{\text{Q},\text{av}} &\coloneqq \frac1{N} \sum_j \sqrt{\sum_{i} \left(d^{\mathrm{II}}_{ij}\right)^2}
    \label{eqn:JQeff}
\end{align}
instead.

\subsection{Extension to describe spin diffusion between two spins}
\label{subsec:results:spindiff}

We consider the spin-diffusion process between a coupled pair of nuclear spins of the same species S. The spins couple through dipolar couplings to an environment of spins, henceforth called bath, of another spin species, I. The bath is assumed to be independent of the two central spins, so that its dynamics can be computed independently. This situation is very typical for low-$\gamma$ spins such as \textsuperscript{13}C embedded in an environment of high-$\gamma$ spins such as \textsuperscript{1}H. We also assume that the geometry of the bath is dense so that each bath spin has a large number of interaction partners and its local dynamics is well captured by spinDMFT. This is usually well justified in three-dimensional crystals. Based on these assumptions, the effect of the bath on the two central spins can be described by two correlated dynamic mean-fields. In addition, the two central spins are also assumed to be subject to a relative chemical shift. A sketch of the effective model is provided in \reffig{fig:model} B.

The mean-field Hamiltonian for this scenario reads
\begin{subequations}
\begin{align}
    \oH(t) &= \oH_{\mathrm{SS}} + \oH_{\text{mf}}(t) + \oH_{\text{CS}}, \\
    \oH_{\mathrm{SS}} &= d^{\text{SS}}_{12} \left(3 \Sz_1 \Sz_2 - \vS_1 \cdot \vS_2\right), \\
    \oH_{\text{mf}}(t) &= \cVz_1(t) \Sz_1 + \cVz_2(t) \Sz_2, \\
    \oH_{\text{CS}} &= \frac{\delta}{2}\left(\Sz_2 - \Sz_1\right),
\end{align}
\label{eqn:Hamiltonian}%
\end{subequations}
where $d^{\text{SS}}_{12}$ is the dipolar coupling between the two S spins and $\delta$ is the difference of the chemical shifts. Since the bath is heteronuclearly coupled to the central spins, only the  $z$-components $\cVz_1$ and $\cVz_2$ of the mean-fields are relevant. Their dynamics, i.e., their second moments can be computed from the spin autocorrelation of the bath spins
\begin{align}
    g_{\text{bath}}^{zz}(t) \coloneqq \expval{\Iz(t) \Iz(0)}
\end{align}
through
\begin{align}
    \mfav{\cVz_i(t)\cVz_j(0)} &= \left(C^{\text{SI}}_{ij}\right)^2 \left(D^{zz}\right)^2 g_{\text{bath}}^{zz}(t), & i,j&\in\{1,2\},
    \label{eqn:spindiffmoments}
\end{align}
where $\dul{D}$ is given by \refeqn{eqn:locenvfields} and we defined the coefficients
\begin{align}
    \left(C^{\text{SI}}_{ij}\right)^2 &\coloneqq \sum_{k} d^{\text{SI}}_{ik} d^{\text{SI}}_{jk}.
\end{align}
Here, $1$ and $2$ are the indices of the two central nuclear spins S. The coefficients are derived from inserting the definition of the local environment fields and neglecting pair correlations, as they are suppressed \cite{graes21}. 
For $i=j$, we obtain a quadratic coupling constant, i.e., 
\begin{align}
    \left(C^{\text{SI}}_{ii}\right)^2 &= \sum_{k} \left(d^{\text{SI}}_{ik}\right)^2 \eqqcolon \left(J^{\text{SI}}_{\text{Q},i}\right)^2.
    \label{eqn:relationCquadcoup}
\end{align}
The Cauchy-Schwarz inequality ensures that the correlation coefficient
\begin{align}
    \rho &\coloneqq \frac{\left(C^{\text{SI}}_{ij}\right)^2}{C^{\text{SI}}_{ii}C^{\text{SI}}_{jj}},
    \label{eqn:relationCrho}
\end{align}
between the two mean-fields is constrained to $\rho\in[-1,1]$ so that the moments are well-defined.

Having defined these expressions, the simulation of the spin-diffusion process requires two steps:
\begin{enumerate}
    \item First, single-site spinDMFT is used to simulate the dynamics and compute the autocorrelation $g_{\text{bath}}^{zz}(t)$ of the bath.
    \item Second, the two central spins are simulated using the Hamiltonian in \refeqn{eqn:Hamiltonian}. The second mean-field moments are determined from \refeqn{eqn:spindiffmoments} using the bath autocorrelation $g_{\text{bath}}^{zz}(t)$ computed in step 1. 
\end{enumerate}
The simulation output of step two can be any quantity that is defined in the Hilbert space of the two central spins. This includes the two-spin correlations 
\begin{align}
    g^{\dira\dirb}_{ij}(t) &\coloneqq \expval{\Sai(t) \Sbj(0)}, & i,j&\in \{1,2\}, & \dira,\dirb&\in \{x,y,z\},
    \label{eqn:autocorrdef}
\end{align}
as well as the zero-quantum (ZQ) correlation function
\begin{align}
    S_{\mathrm{ZQ}}(t) &\coloneqq \frac{\Tr\{ \zqop(t)\zqop(0)\}}{\Tr\{ \zqop(0)^2 \}}, & \zqop&\coloneqq-\frac{\im}{2} \left(\Sp_1\Sm_2 - \Sm_1\Sp_2\right).
    \label{eqn:zqcfunction}
\end{align}
Here, the time evolution of any operator $\op{A}$ is generated by $\op{A}(t)=\oUd \op{A} \oU$ as usual in the Heisenberg picture. For \refeqn{eqn:autocorrdef}, the time evolution is generated by the full spin Hamiltonian, while for \refeqn{eqn:zqcfunction}, the dipolar interaction between $S_1$ and $S_2$ is omitted as in \refcite{suter85}. In total, four different parameters enter the procedure, namely, the bath coupling constant $J^{\text{II}}_{\text{Q,av}}$, which enters directly into the decay of $g_{\text{bath}}^{zz}(t)$, i.e., into the bath correlation time, the pair coupling $d^{\text{SS}}_{12}$, the chemical shift difference $\delta$ and the coefficient tensor $C^{\text{SI}}_{ij}$ containing four elements. All parameters can be computed directly from the geometry of the system.

To test the performance of the spinDMFT description of spin diffusion, we rely on published experimental data of spin diffusion in single crystals. We selected two examples, spin diffusion between the two carbonyl \textsuperscript{13}C atoms in malonic acid \cite{suter85} and spin diffusion between crystallographically distinct \textsuperscript{31}P sites in dipotassium \textalpha-D-glucopyranose-1-phosphate dihydrate (GLP) \cite{kubo88}. Detailed information on the substances and their microscopic structure can be found in Section~\ref{sec:materials}.

\subsection{\textsuperscript{13}C spectral spin diffusion in malonic acid}
\label{subsec:results:malnac}

\begin{figure}
    \centering
    \includegraphics[width=0.5\textwidth]{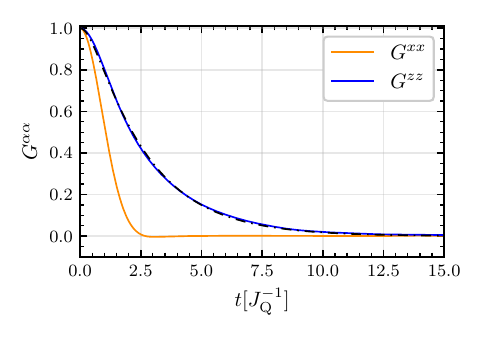}
    \caption{Universal result of the spin autocorrelations simulated by spinDMFT for a homonuclear dipolar coupling. The time axis is given in inverse units of the quadratic coupling constant $J^{\text{HH}}_{\text{Q,av}}$. The black dashed line displays the fit function provided in \refeqn{eqn:spindmft:longitudinalfit}. The transverse correlation $G^{xx}$ has a Gaussian shape  \cite{graes24}.}
    \label{fig:spinDMFT_univ}
\end{figure}

The model described in \refsec{subsec:results:spindiff} is directly applicable to malonic acid identifying the spin species I as \textsuperscript{1}H and S as \textsuperscript{13}C. Due to the molecular geometry and the anisotropy induced by the external field, there are four different categories of proton spins. As an approximation, we consider the bath to be homogeneous using the averaging described in \refeqn{eqn:JQeff} for the quadratic coupling constant. The results for the different crystal orientations are listed in \refsuptab{tab:malnac:JQbath}. In a first step (see the procedure described in \refsec{subsec:results:spindiff}) we simulate the proton-bath dynamics by single-site spinDMFT. The results of the spin autocorrelations for a single homonuclear spin species are depicted in \reffig{fig:spinDMFT_univ}. Note that the time axis is given in units of $1/J^{\text{HH}}_{\text{Q,av}}$ making this graph universal, as the quadratic coupling constant is the only free parameter in single-site spinDMFT. To obtain the results for a given geometry, the time axis needs to be rescaled according to the geometry-specific value of $J^{\text{HH}}_{\text{Q,av}}$. The longitudinal autocorrelation, which is needed to simulate spectral diffusion in the second step, can be approximated well by the fit function \cite{graes24}
\begin{align}
    F^{zz}(t) &= \text{exp}\left(-\gamma \left( \sqrt{t^2 + \kappa^2} - |\kappa| \right)\right),
    & \gamma &\approx 0.43~\JQ,
    & \kappa &\approx 0.65~\JQ^{-1}.
    \label{eqn:spindmft:longitudinalfit}
\end{align}

Several parameters are required to simulate the carbon spectral diffusion in the second step. The carbon dipolar couplings and chemical-shift differences are listed in \refeqn{eqn:malnac:d_and_delta} and the heteronuclear spin mean-field couplings in \refsuptab{tab:malnac:JQCH}. The latter are inserted into \refeqn{eqn:spindiffmoments} together with the longitudinal bath autocorrelation from the first step in order to compute the correlation functions of the mean-fields acting on the carbon spins. These correlation functions define the multivariate Gaussian probability functional of the two mean-fields. In this framework, time propagators are generated by the effective Hamiltonian in \refeqn{eqn:Hamiltonian} for specific time series $\mathcal{V}_1$ and $\mathcal{V}_2$ of the two mean-fields according to 
\begin{align}
    \oU(t)[\mathcal{V}_1,\mathcal{V}_2]&= \mathcal{T} \mathrm{exp}\left( -\texttt{i} \int_0^t \mathrm{d}t' \oH(t')[\mathcal{V}_1,\mathcal{V}_2] \right),
\end{align}
where $\mathcal{T}$ stands for time ordering. Any time-dependent observable is computed by an average over all possible time series weighted by the Gaussian probability functional. This implies for example
\begin{align}
    \mfav{ \expval{\Sx_1(t)\Sx_1(0)} } &= \int \mathrm{D}\mathcal{V}_1 \int \mathrm{D}\mathcal{V}_2 \, p(\mathcal{V}_1,\mathcal{V}_2) \expval{\Sx_1(t)[\mathcal{V}_1,\mathcal{V}_2]\Sx_1(0)},
\end{align}
similar to \refeqn{eqn:mfspinautocorr}. As in single-site spinDMFT, this path integral is numerically evaluated by discretizing the time equidistantly and performing a Monte-Carlo simulation with importance sampling
\begin{align}
    \mfav{ \expval{\Sx_1(t)\Sx_1(0)} }  = \frac1{M} \sum_{i=1}^{M} \expval{\Sx_1(t)[\mathcal{V}_{1,(i)},\mathcal{V}_{2,(i)}]\Sx_1(0)},
\end{align} 
where $M$ is the number of samples, i.e., the number of considered mean-field time series $\mathcal{V}_{1,(i)}$,  $\mathcal{V}_{2,(i)}$. A sampling procedure for this is provided in \refcite{graes21} (Ch. II, Sec. D) and essentially involves a diagonalization of the covariance matrix built from the correlation functions of the mean-fields.

\begin{figure}
    \centering
    \includegraphics[width=0.9\textwidth]{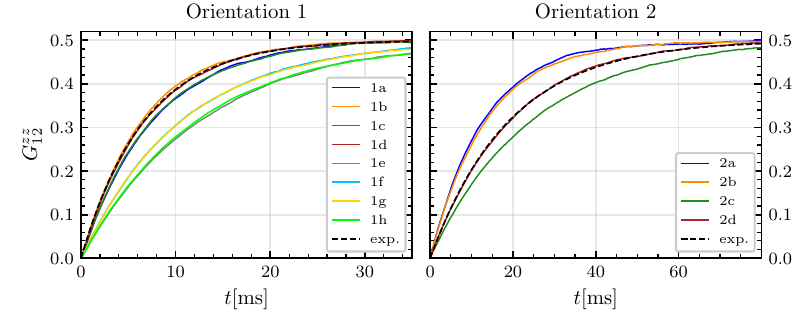}
    \caption{Spectral spin diffusion of carbon spins in malonic acid as seen in the longitudinal pair correlation $G^{zz}_{12}(t)~=~4\expval{\Sz_1(t) \Sz_2(0)}$ for two different crystal orientations. The solid lines correspond to the simulation results by spinDMFT; different possible crystal orientations are shown by different colors. The dashed black line is an exponential fit of the measurement result from \refcite{suter85}. The numerical error of the spinDMFT data is of the order or smaller than 0.02.}
    \label{fig:spindiff_malnac_hom}
\end{figure}

The simulation results for the longitudinal pair correlations are plotted in \reffig{fig:spindiff_malnac_hom} as solid lines for two different crystal orientations measured in experiment. Note that there are several simulation results per experimental crystal orientation because the precise orientation angles $\varphi$ and $\vartheta$ used in experiment are not provided in \refcite{suter85}. As we explain in \refsec{subsec:malnacmaterial}, we infer the angles from the chemical shift and dipolar coupling which is ambiguous and leads to several possibilities, which are all simulated and shown in \reffig{fig:spindiff_malnac_hom}. The resulting rise of the $zz$-correlation from spinDMFT can be fitted well by a mono-exponential function
\begin{align}
    G^{zz}_{12,\text{fit}}(t) &= \frac12 \left(1-\mathrm{e}^{-t/\TSD}\right)
    \label{eqn:spindiff_exponential}
\end{align}
with fit parameters provided in \reftab{tab:malnac:TSDhom}. We stress that we do not assume a mono-exponential behavior, but it occurs in the simulation based on spinDMFT. The black dashed line corresponds to the experimental data. Overall, the agreement is good. All theoretical results decay on the same time scale as the experimental result for both orientations. This should not be taken for granted considering that the bath dynamics occurs on a much shorter time scale on the order of $10-100~\microsecond$. For the first experimental crystal orientation, the orientations 1a-1d agree very well with the experiment, while 1e-1h decay roughly slower by a factor of 1.5-2. For the second experimental crystal orientation, the orientation 2d agrees best, while 2a-2b decay faster and 2c slower than the experimental data. 

\begin{figure}
    \centering
    \includegraphics[width=0.9\textwidth]{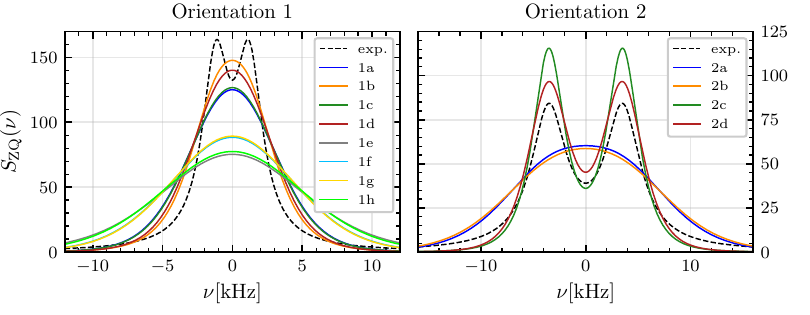}
    \caption{ZQ line of carbon spins in malonic acid for two different crystal orientations. The solid lines correspond to the unbiased prediction by spinDMFT, which is the Fourier transform of  \refeqn{eqn:spinDMFT_ZQC_analytical}; different possible crystal orientations are shown by different colors. The dashed black line is a Lorentzian function with the experimentally determined ZQ line width \cite{suter85}.}
    \label{fig:ZQL_malnac_hom}
\end{figure}

To assess which orientation corresponds best to the one experimentally measured, we compare in addition the ZQ line shapes. In spinDMFT, one can directly simulate the time evolution of the ZQ coherences using \refeqn{eqn:zqcfunction} similar to the spin autocorrelations. However, as we show in \refsupsec{sup:ZQanalytical}, it is also possible to derive an analytical expression for the ZQ time evolution in the framework of spinDMFT. We obtain
\begin{align}
    S_{\mathrm{ZQ}}(t) &= \cos(\delta t) \, \text{exp} \left( -\frac12 \left[\left(J^{\text{CH}}_{\text{Q,}1}\right)^2 + \left(J^{\text{CH}}_{\text{Q,}2}\right)^2 - 2 \rho J^{\text{CH}}_{\text{Q,}1}J^{\text{CH}}_{\text{Q,}2}\right] \int_{0}^{t}\mathrm{d}t_1 \int_{0}^{t}\mathrm{d}t_2 F^{zz}(t_1-t_2) \right),
    \label{eqn:spinDMFT_ZQC_analytical}
\end{align}
where $F^{zz}(t_1-t_2)$ is the fitted bath correlation function provided in \refeqn{eqn:spindmft:longitudinalfit}. The resulting spectrum 
\begin{align}
    S_{\mathrm{ZQ}}(\nu) &= \int_{-\infty}^{\infty} \mathrm{e}^{-2\pi\texttt{i}\nu t} S_{\mathrm{ZQ}}(t) \mathrm{d}t
\end{align}
is compared to the experimentally obtained one in \reffig{fig:ZQL_malnac_hom}. The experimental ZQ line is generated assuming two Lorentzian lines 
\begin{align}
    S^{\text{exp. guess}}_{\mathrm{ZQ}}(\nu) &= \GZQT \left( \frac1{(2\pi\nu-\delta)^2 + \GZQT^2} + \frac1{(2\pi\nu+\delta)^2 + \GZQT^2} \right), & \GZQT &\coloneqq \frac1{\TZQT},
\end{align}
and inserting the measured ZQ time, $\TZQT=\SI{137}{\microsecond}$ for the first and $\TZQT=\SI{78.3}{\microsecond}$ for the second orientation. We emphasize that the assumption of a Lorentzian line shape is very common in NMR, but not necessarily accurate. Indeed, a striking observation in \reffig{fig:ZQL_malnac_hom} is that the unbiased result of spinDMFT deviates considerably from the assumed Lorentzian line shapes for many orientations. Furthermore, the theoretical lines are rather broad, which does not allow the distinction of two separate peaks located at $\pm \delta$, except for orientations 2c-2d. However, a few orientations are reasonably close to the assumed experimental line shape, which, in principle, allows a good match between theory and experiment. Interestingly, the best agreement is obtained for the orientations 1a-1d and 2d. These are the same orientations for which the direct spin-diffusion results agree best, see \reffig{fig:spindiff_malnac_hom} for comparison. This could be an indication that the true experimental orientations are among the best matching ones.

In \refsec{subsec:results:spindiff}, we discussed that the assumption of homogeneity of the bath is valid only if the quadratic coupling constant varies weakly from site to site. As can be seen in \refsuptab{tab:malnac:JQbath}, the relative standard deviation is larger than $1/3$ for a few orientations. This motivated us to perform simulations with an inhomogeneous bath for comparison. The basic idea of this extended approach is to perform an individual simulation of each crystallographically distinct site. Since the distinct sites interact with one another, the self-consistency conditions become intertwined, i.e., the autocorrelation of one site enters the self-consistency condition of another site and vice versa. Therefore, the sites cannot be treated independently of each other. Instead, one has to perform a nested simulation, which solves all self-consistency problems at once. As for the homogeneous case, the resulting converged bath autocorrelations are superposed to obtain the mean-fields acting on the carbon spins. A more thorough formulation of this extension is given in \refsupsec{sup:nestedselfcons}. We stress that the calculation is still a single-site one, but one has to distinguish sites which are not crystallographically equivalent for a given external magnetic field and perform nested spinDMFT simulations. This is similar to the consideration of aluminium nitride via spinDMFT in \refcite{graes26a}.

\begin{figure}
   \centering
   \includegraphics[width=0.9\textwidth]{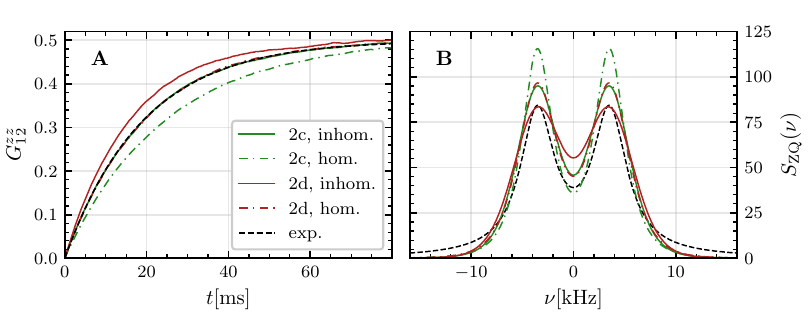}
   \caption{\textbf{Effect of the homogeneous approximation of the proton bath in malonic acid.} The results are shown only for the orientations 2c and 2d, since only in these cases an effect is clearly visible. The dash-dotted lines correspond to the simulation results with homogeneous approximation and the solid lines to the ones resulting from an inhomogeneous bath. The black dashed line corresponds to the experimental result \cite{suter85}. (\textbf{A}) Longitudinal pair correlation $G^{zz}_{12}(t)~=~4\expval{\Sz_1(t) \Sz_2(0)}$. The homogeneous result for 2d, the inhomogeneous result for 2c and the experimental result are barely distinguishable in the plot. (\textbf{B}) ZQ lines. As for the longitudinal pair correlation, the homogeneous result for 2d and the inhomogeneous result for 2c are barely distinguishable. They slightly deviate from the experimental result, but it is not clear whether the assumption of a Lorentzian line shape is correct.}
    \label{fig:malnac_2cd}
\end{figure}

For most orientations, the change in the spin-diffusion result and the ZQ line is marginal when using an inhomogeneous bath, except for 2c-2d, as can be seen in \reffig{fig:malnac_2cd}. The spin-diffusion result for 2c now agrees better with the experiment, while the spin diffusion for orientation 2d is a bit too fast. For both orientations, the deviation of the ZQ line from the Lorentzian line increased at $\nu=0$. This is the relevant frequency for spin diffusion according to 
\begin{align}
    \TSD &= \frac1{d^2 S_{\mathrm{ZQ}}(\nu=0)},
    \label{eqn:TSD_ptbtheory}
\end{align}
which is derived based on the Markov approximation \cite{suter85,Ernst.1998nti}, see also the derivation of the weak-coupling Lindblad formalism \cite{breue06}. The key aspect for this approach to be valid is that the ZQ correlation must die out much faster than the slow diffusion process that we want to capture. Only if this is fulfilled, a mono-exponential correlation ensues. The condition for well-separated time scales needs to be checked carefully before using \eqref{eqn:TSD_ptbtheory}. SpinDMFT and its capability to compute the ZQ line in time is particularly useful in this regard. However, as we mentioned before, the experimental comparison of ZQ lines can only be done on a qualitative level, since the true line shape is not known. The comparison with the inhomogeneous calculation favors orientation 2c, but a clear statement cannot be made as it is difficult to quantify the errors from measurement.

It is worth mentioning that the calculation via the ZQ line (henceforth ZQ formalism) is indeed adequate for malonic acid. This can be clearly seen in \refsuptab{tab:malnac:TSDhom} and \refsuptab{tab:malnac:TSDinhom}, where we list the calculated time constants $\TSD$. The time constants obtained by exponential fits of the direct simulation hardly differ from those obtained by \refeqn{eqn:TSD_ptbtheory}. This observation was also made experimentally, i.e., the measured ZQ time leads to approximately the same $\TSD$ as the one directly measured \cite{suter85}. \emph{A posteriori}, this is not surprising, as the ZQ correlation decays on a time scale of about $10-100~\microsecond$, which is 2-3 orders of magnitude faster than the spin diffusion between the carbon spins. Therefore, the ZQ formalism, which is based on a separation of time scales, is well justified for malonic acid. However, we emphasize that direct calculation of spin diffusion using spinDMFT may also be used in situations where this separation is not given. This is certainly relevant in simulations of spatial spin diffusion, where the bath spins are indistinguishable from the spins between which the polarization transfer occurs. This is typically the case for proton spin diffusion, which is very important in the field of NMR.

\subsection{\textsuperscript{31}P spectral spin diffusion in GLP}
\label{subsec:results:GLP}

Spin diffusion in GLP is conceptually different from that in malonic acid because the phosphor spin ensemble does not separate into isolated spin pairs. To capture spin diffusion between the two crystallographically distinct phosphor sites, we simulate an auxiliary spin pair using an effective coupling
\begin{align}
    \left(d^{\Ph\Ph}_{\eff}\right)^2 &= \sum_{i\in\mathrm{PA}, j\in\mathrm{PB}} \left(d^{\Ph\Ph}_{ij}\right)^2,
\end{align}
which collects the polarization transfer between all different spin pairs $i,j$. Here, PA and PB represent the two categories of $^{31}$P spins. Such an approach is inspired by the perturbation treatment of spin diffusion discussed in the previous section. From \refeqn{eqn:TSD_ptbtheory} it can be deduced that the rate constant is proportional to the square of the coupling scaled by the ZQ line width. Since the rate constants of parallel spin-diffusion pathways are additive, such an effective coupling characterizes the effective rate constant. For the computation of $\left(d^{\Ph\Ph}_{\eff}\right)^2$, we consider all couplings within $10\times10\times10$ unit cells to ensure that finite-size effects are negligible. The computed effective coupling constants can be found in \refcite{datarepo}. The chemical-shift difference between the two $^{31}$P spins is given by
\begin{align}
    \delta(\vartheta) &= \delta_0 \sin(2\vartheta), & \frac{\delta_0}{2\pi} &= \SI{8700}{\hertz}.
\end{align}
In total, there are $30$ different categories of proton spins that interact with the phosphor spin ensemble. As for malonic acid, we first consider the proton bath to be homogeneous. The calculated mean-field parameters can be found in \refcite{datarepo}. In contrast to malonic acid, the mean-field correlation coefficient $\rho$, see \refeqn{eqn:relationCrho}, is very small for GLP, ranging from $0$ to $0.05$. Therefore, its effect on spin diffusion is negligible. In fact, this is required for the approximative treatment using an effective coupling $d_{\eff}$. In general, the spin diffusion process between a pair of spins depends on $\rho$ and must therefore be simulated individually. The situation in GLP is less complicated, as each phosphor spin interacts essentially with a separate local proton environment. The reason for this is that the minimum distance between the phosphor spins is large, so that the local environments  overlap only slightly. This is in contrast to malonic acid, where the minimum distance of the carbon spins is much smaller leading to a considerable correlation coefficient.

\begin{figure}
    \centering
    \includegraphics[width=0.9\textwidth]{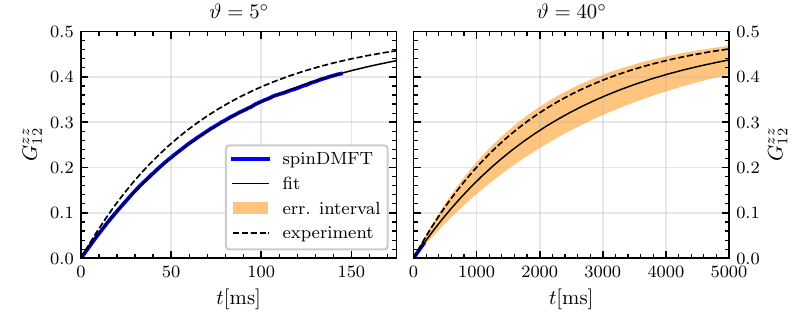}
    \caption{Spectral spin diffusion of $^{31}$P spins in GLP as seen in the longitudinal pair correlation $G^{zz}_{12}(t)~=~4\expval{\Sz_1(t) \Sz_2(0)}$ for two different crystal orientations, $\vartheta=\SI{5}{\degree}$ and $\vartheta=\SI{40}{\degree}$. The simulation is performed until $t\approx\SI{150}{\millisecond}$ using $L=10000$ time steps. The simulation data are shown by the blue line. The black solid line corresponds to an exponential fit according to \refeqn{eqn:spindiff_exponential}. The transparent orange interval indicates the error resulting from the finite time discretization. It is negligible at $\vartheta=\SI{5}{\degree}$, but becomes visible at $\vartheta=\SI{40}{\degree}$, where spin diffusion is considerably slower. The dashed line corresponds to the experimental result \cite{kubo88}.}
    \label{fig:spindiff_GLP_hom_example}
\end{figure}

The simulation results of the spin diffusion process are shown for two different orientations of the crystal in \reffig{fig:spindiff_GLP_hom_example}. For rotation angles around $\vartheta=\SI{45}{\degree}$, spin diffusion is substantially slowed by about two orders of magnitude due to the much larger chemical-shift difference. This complicates the direct calculation by spinDMFT, since the spin diffusion process needs to be captured by the simulation time window and, concurrently, the rapid evolution of the proton spin bath needs to be resolved by the time discretization. This combination of requirements implies a large number of time steps, which is expensive in terms of required memory and computation time. By Fourier transformation and making use of time translation invariance, one can derive a more efficient mean-field sampling procedure, which overcomes this issue, but this is beyond the scope of the present article. Because the time scales of the bath dynamics and the spin diffusion are so well separated, the ZQ formalism is well justified, making the spin diffusion perfectly exponential. This is confirmed by our simulations for all crystal orientations and all different numerical time steps $\delta t$. Consequently, it is not necessary to simulate the spin diffusion process until equilibrium is reached. Instead, we simulate only a small time window with very fine resolution to reduce numerical errors and apply exponential fits as in \refeqn{eqn:spindiff_exponential}. The latter are shown in addition to the simulation results in \reffig{fig:spindiff_GLP_hom_example}. The rising $zz$-correlations corresponding to the experimental time constants are also plotted. The agreement between spinDMFT and the experiment is very good for both examples.

\begin{figure}
    \centering
    \includegraphics[width=0.9\textwidth]{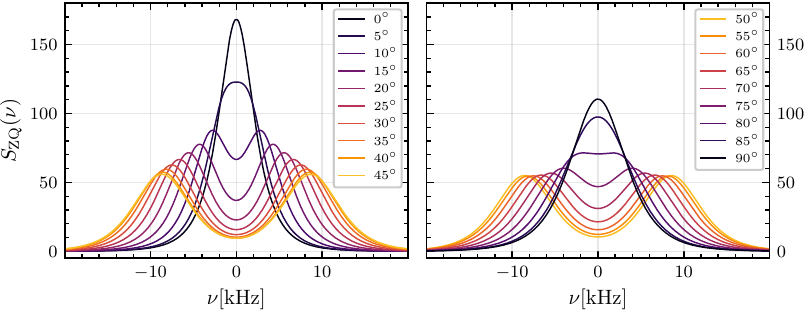}
    \caption{Simulated ZQ line of $^{31}$P spins in GLP for different crystal orientations.}
    \label{fig:GLP_ZQL_phi=0}
\end{figure}

\begin{figure}
    \centering
    \includegraphics[width=0.6\textwidth]{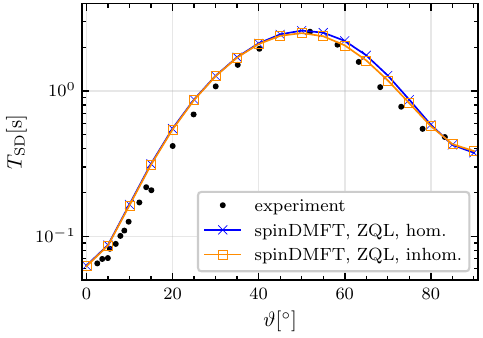}
    \caption{Spin diffusion time constants of GLP as a function of the crystal orientation angle $\vartheta$. The theory results were calculated using the ZQ formalism, since the time scales of spin diffusion and bath dynamics are well separated. The experimental data have been taken from \refcite{kubo88}.}
    \label{fig:GLP_TSD_phi=0}
\end{figure}

As mentioned before, the ZQ formalism is expected to be perfectly justified for GLP as the time scales of the bath dynamics and spin diffusion are well separated: The proton spins interact on a time scale of about $10-\SI{100}{\microsecond}$, while the polarization transfer between the phosphor spins occurs on a time scale of at least $10-\SI{100}{\millisecond}$. The ZQ correlation function can be computed through \refeqn{eqn:spinDMFT_ZQC_analytical} using the proton bath correlation from spinDMFT provided in \refeqn{eqn:spindmft:longitudinalfit} with the appropriate rescaling factor $1/J^{\text{HH}}_{\text{Q,av}}$. The resulting spectra are shown in \reffig{fig:GLP_ZQL_phi=0}. Extracting the value at $\nu=0$ and inserting it in \refeqn{eqn:TSD_ptbtheory} leads to the spin diffusion times plotted in \reffig{fig:GLP_TSD_phi=0}. The agreement between theory and experiment is very good over the whole range of crystal orientations. This is remarkable considering that the time scales cover two orders of magnitude. As in the calculations for malonic acid, this confirms that spinDMFT is an excellent framework for describing spectral spin diffusion. In addition, the spin diffusion times based on direct simulations are identical up to numerical errors, which can be seen in \refsupfig{fig:GLP_TSD_phi=0_with_simulation}. This justifies the use of the ZQ formalism for GLP \emph{a posteriori}. It is worth mentioning that the plane of rotation of the GLP crystal in the experiment is unambiguously specified, but the direction of rotation is not. Hence, there are again two possible orientations for each angle $\vartheta$ (except $\vartheta=\SI{0}{\degree}$ and $\vartheta=\SI{90}{\degree}$). However, we verified that switching the direction of rotation only marginally changes the results.

The proton-proton mean-field coupling varies strongly from site to site for some crystal orientations. Therefore, we repeat the above procedure taking into account the inhomogeneity of the bath. This slightly improves the agreement with the experiment, although the change is very small as can be seen from the two hardly distinguishable curves in \reffig{fig:GLP_TSD_phi=0}.


\section{Discussion}
\label{sec:discussion}

We demonstrated that spinDMFT is an unbiased and accurate approach to the dynamics of disordered spin systems. It is well justified if each spin interacts with a large number of other spins, a situation that is fundamentally important in the field of NMR as it occurs in many samples containing nuclear spins. We emphasize that the calculation through spinDMFT is a first principle one because the dipolar interactions are known once the atomic positions, i.e. the crystal structure, are known. We showed for two examples, malonic acid and GLP, that spectral spin diffusion can readily be simulated. This can be done directly by simulating the polarization transfer from one spin to the other or, quite often, also via the zero-quantum correlation if its dynamics is much faster than the spectral spin diffusion itself. We stress that it is a special asset of spinDMFT to be able to address the zero-quantum correlation without further assumptions, such as uncorrelated single-quantum lines \cite{tycko92}. We also emphasize that the prerequisites for the zero-quantum formalism are optional for spinDMFT, i.e., dynamic mean-fields can be employed also in setups, where perturbation theory breaks down, for example, in homonuclear spatial spin diffusion \cite{suter85}.

Practically, single-site spinDMFT requires the local spin autocorrelations to solve a self-consistency condition. They determine the time correlation of the mean-field, which underlies a Gaussian distribution. Fortunately, the iteration of the self consistency converges quickly within about 5 steps so that the approach is highly efficient computationally. This makes for example calculations of nested spinDMFT easily possible without the need for high-performance computing resources. We demonstrated this by repeating the computations for both test substances taking the inhomogeneity of the respective proton spin ensemble into account. To give an example, a nested spinDMFT calculation for GLP for a single geometry ($30$ sites,  $10^3$ time steps, $2\cdot 10^4$ samples) takes about 5 minutes on a MacBook Pro with Apple M5 chip using 4 cores in parallel.

Once the bath correlations are known, the simulation of spectral spin diffusion is done in a further step, namely the simulation of the involved spins under the bath mean-fields. This step is computationally cheap as long as the time scale of spin diffusion and bath dynamics are not too strongly separated. If the time scales are well separated, the zero-quantum formalism is well justified allowing an efficient alternative, which we also demonstrated for both test substances. We recommend the use of the analytical equation for the zero-quantum correlation, \refeqn{eqn:spinDMFT_ZQC_analytical}. The evaluation of this formula only requires the computation of coupling sums, which are easily obtained from the system geometry, and a rescaling of the universal longitudinal bath correlation from single-site spinDMFT.

A limitation of spinDMFT is the requirement for large coordination numbers, which excludes for example one-dimensional systems or systems where each spin has only a single dominant interaction partner. If the system under study can be partitioned into small groups of strongly interacting spins, cluster spinDMFT (CspinDMFT) \cite{graes23} can help out at the cost of increasing the computational effort. We would like to emphasize that the presence of long-range interactions is very advantageous for spinDMFT because it generally increases the effective coordination number justifying the approximation even better. In this regard, spinDMFT is complementary to many brute-force methods such as exact diagonalization, which are suitable rather for low-dimensional, less connected systems.

As can be seen in the example of GLP, spin diffusion can depend very strongly on the orientation of the crystal. The orientation modifies not only the strength of the coupling between the two spins but also the bath dynamics, which is reflected in the coupling constants in spinDMFT. The approach can be straightforwardly extended to powder samples by introducing an additional average of the time-evolved expectation values over all possible crystal orientations. In this way, also disordered systems like frozen glasses or amorphous samples are accessible as long as their geometries can be deduced.

SpinDMFT can and needs to be extended in various ways to be applicable to a broader class of magnetic resonance experiments. An important technique to be considered in future studies is magic-angle spinning, which is applied in most high-resolution solid-state NMR experiments. A promising outlook of this article is the description of spatial spin diffusion as it occurs in dynamic nuclear polarization and generally between proton spins. The sheer number of spins involved in such scenarios excludes most brute-force descriptions and demands an effective model such as the one imposed by spinDMFT. Besides this, it is also interesting to extend spinDMFT to finite temperatures, which is currently pursued.

\section{Materials and Methods}
\label{sec:materials}

\begin{figure}
    \centering
    \includegraphics[width=0.9\textwidth]{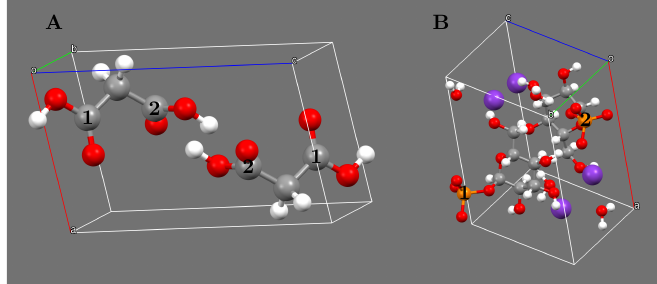}
    \caption{\textbf{Unit cells of the considered test substances.} White, grey, red, purple and orange spheres correspond to hydrogen, carbon, oxygen, potassium and phosphor atoms, respectively. (\textbf{A}) Malonic acid. The unit cell contains four \textsuperscript{13}C spins, which are labelled here by black numbers. Due to inversion symmetry, two carbon pairs are each equivalent in terms of dipolar couplings and chemical shifts and therefore obtain the same number. (\textbf{B}) GLP. The unit cell contains two crystallographically distinct \textsuperscript{31}P spins, which are also labelled by black numbers.}
    \label{fig:unitcell}
\end{figure}

The simulation of spin diffusion in the two test substances, malonic acid and dipotassium \textalpha-D-glucopyranose-1-phosphate dihydrate (GLP), by spinDMFT requires the respective crystal geometry as input. The unit cells of the two crystals are shown in \reffig{fig:unitcell}.

\subsection{Malonic acid}
\label{subsec:malnacmaterial}

For malonic acid (propanedioic acid, C$_3$H$_4$O$_4$), two single crystal spin-diffusion measurements and ZQ line shapes of the \textsuperscript{13}C carbonyl resonances have been published.  Unfortunately, the orientations of the crystal relative to the magnetic field are not clearly indicated in the publications \cite{suter85,Suter.1985a}. However, at least the values of the chemical-shift differences and the magnitude of the dipolar couplings are provided,
\begin{subequations}
\label{eqn:malnac:d_and_delta}
\begin{align}
    \text{Orientation 1:} & & \frac{d^{\text{CC}}_{12}}{2\pi} &= \SI{230}{\hertz}, & \frac{\delta}{2\pi} &= \SI{1200}{\hertz}, \\
    \text{Orientation 2:} & & \frac{d^{\text{CC}}_{12}}{2\pi} &= \SI{250}{\hertz}, & \frac{\delta}{2\pi} &= \SI{3530}{\hertz},
\end{align}
\end{subequations}
which allows us to reconstruct the orientations based on the published crystal structure \cite{Jagannathan.1994}. The intramolecular distance of the two carbonyl atoms is 2.502 {\AA} while the shortest intermolecular distance between two non-equivalent carbonyls is 6.0 {\AA}. Due to the large difference in distances, we restricted the calculation to intramolecular spin diffusion since the rate constant to the closest intermolecular neighbor would be smaller by a factor of 200.
To determine the dipolar coupling and the chemical shifts for a given orientation, the crystal structure was geometry optimized in CASTEP \cite{CASTEP} using a density functional PBE \cite{PBE} with a cutoff energy of 900 eV. In a second run, chemical-shielding parameters were calculated using the same density functional \cite{GIPAW,GIPAW_USP,magres_review}. The dipolar-coupling values were calculated based on the geometry-optimized crystal structure. The distance of the two carbonyl \textsuperscript{13}C atoms was 2.502 {\AA} corresponding to an anisotropy of the dipolar-coupling of $\delta_{12}^\mathrm{CC}/{2\pi} = -2\frac{\mu_0}{4\pi^2}\frac{\gamma_\mathrm{C}^2\hbar}{r_\mathrm{CC}^3} = \SI{-970}{\hertz}$. The orientation of the internuclear C-C vector with respect to the static magnetic field is described by the two Euler angles $\vartheta$ and $\varphi$. The angle $\vartheta$ can be inferred from the experimentally given dipolar coupling using $d_{12}^\mathrm{CC} = \delta_{12}^\mathrm{CC} \mathrm{P}_2(\cos\vartheta)/2$ while the angle $\varphi$ can be inferred from the chemical-shift difference by matching the chemical-shift differences calculated using DFT to the ones provided experimentally for the two orientations. This procedure leads to a total of 24 possible orientations, where always two of them are symmetry related and give the same values for the sum of the squared couplings. This leaves us with eight possible solutions for orientation 1 and four possible solutions for orientation 2 \cite{suter85}. The Euler angles are summarized in \reftab{tab:malnac:orientations}. 


\begin{table}[h]
    \centering
    \caption{Possible orientations of the CC bond in malonic acid relative to the static field.}
    \label{tab:malnac:orientations}
    \begin{tabular}{c|c|c|c}
        \hline
        Orientation & $\vartheta$ & $\varphi$ & label\\
        \hline
         1 & $36.3^\circ$ & $114.7^\circ$ & (1a)\\
           &              & $164.1^\circ$ & (1b)\\
           &              & $178.6^\circ$ & (1c)\\
           &              & $223.9^\circ$ & (1d)\\
           & $82.5^\circ$ &  $54.7^\circ$ & (1e)\\
           &              & $111.3^\circ$ & (1f)\\
           &              & $240.0^\circ$ & (1g)\\
           &              & $304.9^\circ$ & (1h)\\
           \hline
         2 & $34.6^\circ$ &  $74.5^\circ$ & (2a)\\
           &              & $252.3^\circ$ & (2b)\\
           & $90.0^\circ$ &  $33.8^\circ$ & (2c)\\
           &              & $141.6^\circ$ & (2c)\\           
        \hline
    \end{tabular}
\end{table}

\subsection{Dipotassium \textalpha-D-glucopyranose-1-phosphate dihydrate}
\label{subsec:GLPmaterial}

For dipotassium \textalpha-D-glucopyranose-1-phosphate dihydrate (C$_6$H$_{11}$K$_2$O$_9$P$\cdot$2H$_2$O, GLP) spin-diffusion data as a function of crystal orientation are available in the literature \cite{kubo88}. The situation here is different from the case of malonic acid since only a single \textsuperscript{31}P atom is contained in GLP but there are two crystallographically distinct sites with different orientation-dependent chemical shifts. The crystal structure of GLP \cite{Beevers.1965} was geometry optimized in CASTEP \cite{CASTEP} using a density functional PBE \cite{PBE} with a cutoff energy of 900 eV. In a second run, chemical-shielding parameters were calculated using the same density functional \cite{GIPAW,GIPAW_USP,magres_review}. The dipolar-coupling values were calculated based on the geometry-optimized crystal structure. The chemical-shift difference of the two distinct sites was calculated using the obtained chemical-shift tensors as a function of the crystal orientation. There are no well separated pairs of \textsuperscript{31}P spins in the sample, requiring us to take into account more than two phosphor spins. We addressed this issue in the beginning of \refsec{subsec:results:GLP}.


\clearpage 

%
\bibliography{sciadv_literature} 
\bibliographystyle{sciencemag}

%
%
%
%
%
%


\section*{Acknowledgments}

We are thankful to Luca J. Oelkrug, Przemyslav Bieniek, Antonia J. Bock and Joachim Stolze for useful discussions.

\paragraph*{Funding:}
GSU acknowledges funding by the German Science Foundation (DFG) in project UH~90/14-2. ME acknowledges funding by ETH Zürich.

\paragraph*{Author contributions:}
All authors conceived the fundamental idea of applying spinDMFT to spectral spin diffusion. ME conducted preliminary calculations and analyzed the experimental data including the reconstruction of the crystal orientations. TG performed the simulations and created all figures. All authors participated in interpreting the simulated data and in writing and editing the manuscript.

\paragraph*{Competing interests:}
There are no competing interests to declare.

\paragraph*{Data and materials availability:}
An open-source code collection for spinDMFT has been published under \refcite{graes25}. Data regarding the system geometries and simulation results can be found in \refcite{datarepo}.

\subsection*{Supplementary materials}
Section S1 Extension to an inhomogeneous bath\\
Section S2 Derivation of an analytical expression for the ZQ line\\
Section S3 Coupling constants and diffusion times for spin diffusion in malonic acid\\
Section S4 Additional plots for spin diffusion in GLP\\
Figure S1\\
Tables S1 to S4\\


\newpage


\renewcommand{\thefigure}{S\arabic{figure}}
\renewcommand{\thetable}{S\arabic{table}}
\renewcommand{\theequation}{S\arabic{equation}}
\renewcommand{\thepage}{S\arabic{page}}
\setcounter{figure}{0}
\setcounter{table}{0}
\setcounter{equation}{0}
\setcounter{page}{1} 


\begin{center}
\section*{Supplementary Materials for\\ \scititle}
Timo~Gräßer$^{\ast}$,
G\"otz~S.~Uhrig,
Matthias~Ernst$^{\ast}$\and

\small $^\ast$Corresponding authors. Email: timo.graesser@phys.chem.ethz.ch, maer@ethz.ch
\end{center}

\renewcommand{\thesection}{S\arabic{section}}
\renewcommand{\thesubsection}{S\arabic{section}.\arabic{subsection}}
\setcounter{section}{0}

\subsubsection*{This PDF file includes:}
Section S1 Extension to an inhomogeneous bath\\
Section S2 Derivation of an analytical expression for the ZQ line\\
Section S3 Coupling constants and diffusion times for spin diffusion in  malonic acid\\
Section S4 Additional plots for spin diffusion in GLP\\
Figure S1\\
Tables S1 to S4\\
\newpage





\section{Extension to an inhomogeneous bath}
\label{sup:nestedselfcons}

In this section, we extend the effective models presented in the main text to systems with spatial inhomogeneity that are still dense enough for the mean-field approximation to be justified. We consider a bath containing $N_0$ different categories of spins. In this context, the word category does not refer to the species of the nucleus, but to the surrounding geometry, i.e., we still consider the bath to be homonuclear containing spins of the same species I. The spins may interact not only within but also among the different categories so that it is not valid to solve the self-consistency problem separately for each bath-spin category. Instead, one has to use either cluster spinDMFT (CspinDMFT \cite{graes23}), which can entail large computational effort and is only required in case of small coordination numbers, or an adapted, nested self-consistency calculation. 

The latter is set up as follows: For each category $K\in\{1,\dots,N_0\}$, a spin on a representative site $k$ is simulated separately in its own mean-field according to 
\begin{align}
    \mb{H}^{\text{mf}}_{k} &= \vec{V}_{k}(t) \cdot \vS_{k}.
\end{align}
The required second mean-field moments are calculated by the self-consistency condition 
\begin{subequations}
\begin{align}
    \mfav{V_k^{\dira}(t)V_k^{\dira}(0)} &= \left(D^{\dira\dira}\right)^2 \sum_{m,n} d_{km}^{\text{II}} d_{kn}^{\text{II}} \expval{\Sa_m(t)\Sa_n(0)} \\
    &= \left(D^{\dira\dira}\right)^2 \sum_{m} \left(d_{km}^{\text{II}}\right)^2 \, g^{\dira\dira}_{mm}(t) \\
    &= \left(D^{\dira\dira}\right)^2 \sum_{L=1}^{N_0} \left(J^{\text{II}}_{\text{Q},KL}\right)^2 \, g^{\dira\dira}_{L}(t),
    \label{eqn:nestedselfcons}
\end{align}
\end{subequations}
where $g^{\dira\dira}_{L}(t)$ is an autocorrelation of a spin of category $L$. The quadratic coupling sums are given by
\begin{align}
    \left(J^{\text{II}}_{\text{Q},KL}\right)^2 &\coloneqq \sum_{l\in L} \left(d^{\text{II}}_{kl}\right)^2,
\end{align}
i.e., they sum up the squared couplings between the representative spin $k$ of category $K$ and \emph{all} spins $l$ of category $L$. The nested self-consistency problem is solved by simulating the categories in parallel and updating all mean-fields at the end of each iteration step using \refsupeqn{eqn:nestedselfcons}. In contrast to CspinDMFT, this procedure scales only linear with the number $N_0$ of categories and is thus much less costly. In case of small coordination numbers, e.g., if a pair of bath spins is rather close, a hybrid approach which combines CspinDMFT with the nested self-consistency calculation is conceivable.

The presented method yields a set of $N_0$ bath autocorrelations. The mean-field correlations for a second spin sort S interacting with the I-spin bath is given by
\begin{align}
    \mfav{V_i^{z}(t)V_j^{z}(0)} &= \left(D^{zz}\right)^2 \sum_{K=1}^{N_0} \left(C^{\text{SI}}_{ij,K}\right)^2 g^{zz}_{K}(t,0)
    \label{eqn:mfcoupling_inhom}
\end{align}
defining
\begin{align}
    \left(C^{\text{SI}}_{ij,K}\right)^2 &\coloneqq \sum_{k\in K} d^{\text{SI}}_{ik} d^{\text{SI}}_{jk}
\end{align}
for all pairs $i,j$ of spins S. This is derived by inserting the definition of the local environment fields and neglecting pair correlations. The consideration of an inhomogeneous bath  complicates the first step of the simulation of spin diffusion, since the nested self-consistency has to be implemented. However, except for the adapted formula for the mean-field correlations, see \refsupeqn{eqn:mfcoupling_inhom}, the simulation of S spins is essentially the same as before.

\section{Derivation of an analytical expression for the ZQ line}
\label{sup:ZQanalytical}

In this supplementary section, we derive a compact analytical formula for the ZQ correlation function in the framework of spinDMFT. As a starting point, we consider the definition
\begin{align}
    S_{\mathrm{ZQ}}(t) &\coloneqq \frac{\Tr\{ \zqop(t)\zqop(0)\}}{\Tr\{ \zqop(0)^2 \}}, &
    \zqop &\coloneqq -\frac{\im}{2} \left(\Sp_1\Sm_2 - \Sm_1\Sp_2\right) = - \left(\Sx_1\Sy_2 + \Sy_1 \Sx_2\right).  
\end{align}
In spinDMFT, the bath degrees of freedom are captured by the mean-fields, which entails
\begin{subequations}
\begin{align}
    S_{\mathrm{ZQ}}(t) &\approx \int \mathrm{D}\mathcal{V}_1 \int \mathrm{D}\mathcal{V}_2 \, p(\mathcal{V}_1, \mathcal{V}_2) \frac{\Tr_{\loc}\{ \zqop(t)\zqop(0)\}}{\Tr_{\loc}\{ \zqop(0)^2 \}} \\
    &= 16 \int \mathrm{D}\mathcal{V}_1 \int \mathrm{D}\mathcal{V}_2 \, p(\mathcal{V}_1, \mathcal{V}_2) \Bigl(\Tr_{\loc}(\Sx_1(t)\Sx_1(0)) \Tr_{\loc}(\Sx_2(t)\Sx_2(0)) \nonumber\\
    &\qquad\qquad\qquad- \Tr_{\loc}(\Sx_1(t)\Sy_1(0)) \Tr_{\loc}(\Sx_2(t)\Sy_2(0))\Bigr),
\end{align}
\end{subequations}
where in the last step we make use of rotational symmetry around $z$ in spin space. For malonic acid and GLP, the proton bath couples heteronuclearly to the spins of interest. Therefore, time ordering can be omitted and the time propagators and time-evolved spin operators simplify according to
\begin{subequations}
\begin{align}
    \oU(t)[\mathcal{V}_i] &= \exp\left( -\im \left(f_i(t)+\delta_it\right) \Sz_{i} \right) \\
    \Sx_i(t)[\mathcal{V}_i] &= \cos\left(f_i(t)+\delta_it\right) \Sx_i - \sin\left(f_i(t)+\delta_it\right) \Sy_i
\end{align}
\end{subequations}
where 
\begin{align}
    f_i(t) &= \int_{0}^{t} V_i^{z}(t')\mathrm{d}t'
\end{align}
is the accumulated phase resulting from the bath and $\delta_i$ are the individual chemical shift of the two spins under consideration. Inserting $\Sx_i(t)[\mathcal{V}_i]$ and carrying out the trace yields
\begin{align}
    S_{\mathrm{ZQ}}(t) &= 4 \Bigl( \mfav{\cos(f_1(t)+\delta_1t)\cos(f_2(t)+\delta_2t)} - \mfav{\sin(f_1(t)+\delta_1t)\sin(f_2(t)+\delta_2t)} \Bigr),
\end{align}
where we use the overline as a shorthand for the mean-field average. Using Euler's formula and the standard trick
\begin{align}
    \overline{\exp{x}} &= \exp{\frac12 \overline{x^2}}.
\end{align}
for Gaussian probability functionals, we deduce
\begin{align}
    S_{\mathrm{ZQ}}(t) &= \cos(\delta_1t-\delta_2t) \exp{-\frac12 \mfav{\left(f_1(t)-f_2(t)\right)^2}}.
\end{align}
Inserting the accumulated phases, the exponent can be expressed as 
\begin{align}
    \mfav{\left(f_1(t)-f_2(t)\right)^2} &= \int_0^{t} \mathrm{d}t_1\int_0^{t} \mathrm{d}t_2 \mfav{\left(V^{z}_1(t_1)-V^{z}_2(t_1)\right)\left(V^{z}_1(t_2)-V^{z}_2(t_2)\right)}.
\end{align}
In case of a homogeneous bath, we use \refeqn{eqn:spindiffmoments} to derive the final expression
\begin{align}
    S_{\mathrm{ZQ}}(t) &= \cos(\delta t) \, \text{exp} \left( -\frac12 \left[\left(J^{\text{CH}}_{\text{Q,}1}\right)^2 + \left(J^{\text{CH}}_{\text{Q,}2}\right)^2 - 2 \rho J^{\text{CH}}_{\text{Q,}1}J^{\text{CH}}_{\text{Q,}2}\right] \int_{0}^{t}\mathrm{d}t_1 \int_{0}^{t}\mathrm{d}t_2 F^{zz}(t_1-t_2) \right).
\end{align}
In case of an inhomogeneous bath, we analogously use \refsupeqn{eqn:mfcoupling_inhom} to obtain
\begin{align}
    S_{\mathrm{ZQ}}(t) &= \cos(\delta t) \, \text{exp} \left( -\frac12 \sum_{K=1}^{N_0} \left(C^{\text{SI}}_{ij,K}\right)^2 \left(2\delta_{ij}-1\right) \int_{0}^{t}\mathrm{d}t_1 \int_{0}^{t}\mathrm{d}t_2 G^{zz}_{K}(t_1-t_2) \right),
    \label{eqn:spinDMFT_ZQC_analytical_2}
\end{align}
where $G^{zz}_{K}(t)$ is the normalized bath correlation for bath spin category $K$.

\section{Coupling constants and diffusion times for spin diffusion in  malonic acid}
\label{sup:malnac_constants}

The tables \ref{tab:malnac:JQbath} and \ref{tab:malnac:JQCH} contain the coupling constants obtained from the crystal geometry of malonic acid. They are required as input parameters for the simulations by spinDMFT presented in \refsec{subsec:results:malnac}. The tables 
\ref{tab:malnac:TSDhom} and \ref{tab:malnac:TSDinhom} contain the time constants for $^{13}$C spectral spin diffusion in malonic acid extracted from fitting and calculated through the ZQ formalism on the basis of spinDMFT.

\begin{table}[h]
    \centering
    \caption{Averages and relative standard deviations of the bath coupling constants for $^{1}$H spins in malonic acid for possible crystal orientations. The averages $\mu$ are computed by averaging $J_\mathrm{Q,i}^{\mathrm{HH}}=\sqrt{\sum_j \left(d^{\mathrm{HH}}_{ij}\right)^2}$ with respect to the sites $i$ as in \refeqn{eqn:JQeff}. The relative standard deviations are given by $\sigma_{\mathrm{rel}} = \sqrt{\overline{x^2}-\mu^2}/\mu$ with $x=J_\mathrm{Q,i}^{\mathrm{HH}}$.}
    \label{tab:malnac:JQbath}
\begin{tabular}{c|cc|cc}
    \hline
    & \multicolumn{2}{c|}{Ori. 1} & \multicolumn{2}{c}{Ori. 2} \\
    Subori. & $\mu\left(J_\mathrm{Q,i}^{\mathrm{HH}}\right)/2\pi$ & $\sigma_{\mathrm{rel}}\left(J_\mathrm{Q,i}^{\mathrm{HH}}\right)$ & $\mu\left(J_\mathrm{Q,i}^{\mathrm{HH}}\right)/2\pi$ & $\sigma_{\mathrm{rel}}\left(J_\mathrm{Q,i}^{\mathrm{HH}}\right)$ \\
    \hline
    a & \SI{4509}{\hertz} & 0.18 & \SI{8583}{\hertz} & 0.282 \\
    b & \SI{6993}{\hertz} & 0.135 & \SI{7732}{\hertz} & 0.358 \\
    c & \SI{4717}{\hertz} & 0.203 & \SI{9238}{\hertz} & 0.336 \\
    d & \SI{5258}{\hertz} & 0.318 & \SI{6879}{\hertz} & 0.415 \\
    e & \SI{7849}{\hertz} & 0.239 & - & - \\
    f & \SI{4418}{\hertz} & 0.243 & - & - \\
    g & \SI{4422}{\hertz} & 0.19 & - & - \\
    h & \SI{5222}{\hertz} & 0.185 & - & - \\
    \hline
    \end{tabular}
\end{table}

\newpage

\begin{table}[h]
    \centering
    \caption{Magnitudes $J_{\text{Q},i}^{\text{CH}}$ and correlation coefficients $\rho$ of the mean-fields for \textsuperscript{13}C spins in malonic acid for possible crystal orientations. The parameters are related to the coefficients $C_{ij}^{\mathrm{CH}}$ required for the self-consistency condition in \refeqn{eqn:spindiffmoments} according to \refeqn{eqn:relationCquadcoup} and \refeqn{eqn:relationCrho}.}
    \label{tab:malnac:JQCH}
    \begin{tabular}{c|ccc|ccc}
    \hline
    & \multicolumn{3}{c|}{Ori. 1} & \multicolumn{3}{c}{Ori. 2} \\
    Subori. & $J_\mathrm{Q,1}^{\mathrm{CH}}/2\pi$ & $J_\mathrm{Q,2}^{\mathrm{CH}}/2\pi$ & $\rho$ & $J_\mathrm{Q,1}^{\mathrm{CH}}/2\pi$ & $J_\mathrm{Q,2}^{\mathrm{CH}}/2\pi$ & $\rho$ \\
    \hline
    a & \SI{3240}{\hertz} & \SI{2720}{\hertz} & 0.45 & \SI{3585}{\hertz} & \SI{2675}{\hertz} & -0.48 \\
    b & \SI{3180}{\hertz} & \SI{1585}{\hertz} & 0.5 & \SI{3617}{\hertz} & \SI{3371}{\hertz} & -0.33 \\
    c & \SI{2717}{\hertz} & \SI{3053}{\hertz} & 0.42 & \SI{2878}{\hertz} & \SI{2665}{\hertz} & 0.58 \\
    d & \SI{2440}{\hertz} & \SI{3019}{\hertz} & 0.48 & \SI{2971}{\hertz} & \SI{2624}{\hertz} & 0.55 \\
    e & \SI{3686}{\hertz} & \SI{3021}{\hertz} & -0.39 & - & - & - \\
    f & \SI{3164}{\hertz} & \SI{2663}{\hertz} & -0.2 & - & - & - \\
    g & \SI{3135}{\hertz} & \SI{2718}{\hertz} & -0.16 & - & - & - \\
    h & \SI{3461}{\hertz} & \SI{3152}{\hertz} & -0.25 & - & - & - \\
    \hline
    \end{tabular} 
\end{table}

\newpage

\begin{table}[h]
    \centering
    \caption{Time constants $\TSD$ for carbon spin diffusion in malonic acid for different crystal orientations. The term \enquote{direct} refers to the data extracted from the spinDMFT simulation according to the exponential fit function in \refeqn{eqn:spindiff_exponential}. The provided errors result from the time discretization, which is the dominant source of numerical error. The fitting errors are considerably smaller. The term \enquote{ZQ} refers to the data extracted from perturbation theory using the calculated ZQ line. The experimental results are taken from \refcite{suter85}.}
    \label{tab:malnac:TSDhom}
    \begin{tabular}{c|cc|cc}
    \hline
    & \multicolumn{2}{c|}{Ori. 1} & \multicolumn{2}{c}{Ori. 2} \\
    \hline
    label & direct & ZQL & direct & ZQL \\
    \hline
    a & \SI{7.56$\pm$0.05}{\millisecond} & \SI{7.65}{\millisecond} & \SI{12.91$\pm$0.31}{\millisecond} & \SI{13.45}{\millisecond} \\
    b & \SI{6.49$\pm$0.12}{\millisecond} & \SI{6.48}{\millisecond} & \SI{13.52$\pm$0.7}{\millisecond} & \SI{13.81}{\millisecond} \\
    c & \SI{7.51$\pm$0.07}{\millisecond} & \SI{7.56}{\millisecond} & \SI{24.44$\pm$1.55}{\millisecond} & \SI{22.47}{\millisecond} \\
    d & \SI{6.78$\pm$0.04}{\millisecond} & \SI{6.83}{\millisecond} & \SI{19.14$\pm$0.9}{\millisecond} & \SI{17.93}{\millisecond} \\
    e & \SI{12.51$\pm$0.49}{\millisecond} & \SI{12.72}{\millisecond} & - & - \\
    f & \SI{10.65$\pm$0.01}{\millisecond} & \SI{10.83}{\millisecond} & - & - \\
    g & \SI{10.71$\pm$0.16}{\millisecond} & \SI{10.72}{\millisecond} & - & - \\
    h & \SI{12.26$\pm$0.23}{\millisecond} & \SI{12.36}{\millisecond} & - & - \\
    \hline
    exp. & \SI{6.7}{\millisecond} & \SI{7.2}{\millisecond} & \SI{19.15}{\millisecond} & \SI{20.8}{\millisecond} \\
    \hline
    \end{tabular}
\end{table}

\newpage

\begin{table}[h]
    \centering
    \caption{Same as \refsuptab{tab:malnac:TSDhom} but with an inhomogeneous proton bath considered in the simulations.}
    \label{tab:malnac:TSDinhom}
    \begin{tabular}{c|cc|cc}
    \hline
    & \multicolumn{2}{c|}{Ori. 1} & \multicolumn{2}{c}{Ori. 2} \\
    \hline
    label & direct & ZQL & direct & ZQL \\
    \hline
    a & \SI{7.5$\pm$0.02}{\millisecond} & \SI{7.61}{\millisecond} & \SI{12.9$\pm$0.34}{\millisecond} & \SI{13.44}{\millisecond} \\
    b & \SI{6.41$\pm$0.04}{\millisecond} & \SI{6.47}{\millisecond} & \SI{13.36$\pm$0.59}{\millisecond} & \SI{13.77}{\millisecond} \\
    c & \SI{7.09$\pm$0.43}{\millisecond} & \SI{7.52}{\millisecond} & \SI{19.2$\pm$0.44}{\millisecond} & \SI{17.68}{\millisecond} \\
    d & \SI{6.81$\pm$0.08}{\millisecond} & \SI{6.83}{\millisecond} & \SI{15.76$\pm$0.24}{\millisecond} & \SI{14.64}{\millisecond} \\
    e & \SI{12.5$\pm$0.59}{\millisecond} & \SI{12.6}{\millisecond} & - & - \\
    f & \SI{10.79$\pm$0.06}{\millisecond} & \SI{10.87}{\millisecond} & - & - \\
    g & \SI{10.71$\pm$0.36}{\millisecond} & \SI{10.75}{\millisecond} & - & - \\
    h & \SI{12.17$\pm$0.18}{\millisecond} & \SI{12.35}{\millisecond} & - & - \\
    \hline
    exp. & \SI{6.7}{\millisecond} & \SI{7.2}{\millisecond} & \SI{19.15}{\millisecond} & \SI{20.8}{\millisecond} \\
    \hline
    \end{tabular}
\end{table}

\newpage

\section{Additional plots for spin diffusion in GLP}
\label{sup:GLP_plots}

\begin{figure}[h]
    \centering
    \includegraphics[width=0.6\textwidth]{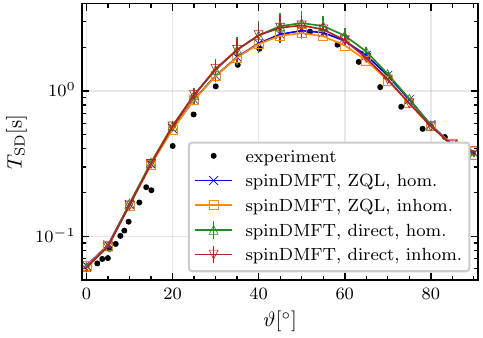}
    \caption{Same as \reffig{fig:GLP_TSD_phi=0}, but with the directly simulated spinDMFT data included. The error bars correspond to the estimated error interval due the time discretization error.}
    \label{fig:GLP_TSD_phi=0_with_simulation}
\end{figure}

\reffig{fig:GLP_TSD_phi=0_with_simulation} contains the direct simulation results for spectral spin diffusion of $^{31}$P spins in GLP, which is considered in \refsec{subsec:results:GLP}. For the simulation input parameters, we direct the reader to the data repository in \refcite{datarepo}.

\end{document}